# Molecular Beam Epitaxy of GaN Nanowires on Epitaxial Graphene

Sergio Fernández-Garrido,*,† Manfred Ramsteiner,† Guanhui Gao,† Lauren A.Galves,† Bharat Sharma,† Pierre Corfdir,† Gabriele Calabrese,† Ziani de Souza Schiaber,‡,¶ Carsten Pfüller,† Achim Trampert,† João Marcelo J. Lopes,† Oliver Brandt,† and Lutz Geelhaar†

†Paul-Drude-Institut für Festkörperelektronik, Hausvogteiplatz 5–7, 10117 Berlin, Germany
‡Laboratório de Filmes Semicondutores Universidade Estadual Paulista Bauru, 17033-360 São Paulo, Brazil
¶On leave at Paul-Drude-Institut für Festkörperelektronik from her home institution

E-mail: garrido@pdi-berlin.de

## Abstract

We demonstrate an all-epitaxial and scalable growth approach to fabricate single-crystalline GaN nanowires on graphene by plasma-assisted molecular beam epitaxy. As substrate, we explore several types of epitaxial graphene layer structures synthesized on SiC. The different structures differ mainly in their total number of graphene layers. Because graphene is found to be etched under active N exposure, the direct growth of GaN nanowires on graphene is only achieved on multilayer graphene structures. The analysis of the nanowire ensembles prepared on multilayer graphene by Raman spectroscopy and transmission electron microscopy reveals the presence of graphene underneath as well as in between nanowires, as desired for the use of this material as contact layer in nanowire-based devices. The nanowires nucleate preferentially at step edges, are vertical, well aligned, epitaxial, and of comparable structural quality as similar structures fabricated on conventional substrates.

*Keywords*: singlelayer graphene, bilayer graphene, multilayer graphene, nanocolumn, nanorod, III-V compound semiconductors, nitrides, growth

Graphene offers new opportunities to improve the performance and functionalities of optoelectronic devices based on III-V compound semiconductors. The transparency and high in-plane electrical conductivity of graphene and few-layer graphene films is of interest for the fabrication of contact layers in light emitting diodes (LEDs), solar cells, and photodetectors, especially in the deep-UV spectral region where standard ITO contacts are not transparent.[1–8] At the same time, its high thermal conductivity is potentially helpful to dissipate heat, a major issue when light emitters are operated at high currents. Furthermore, provided that III-V compounds could be synthesized with high structural perfection on graphene, the flexibility of this material as well as the possibility to be grown or transferred onto different substrates may enable the integration of III-V based devices with otherwise incompatible material systems such as the complementary metal-oxide-semiconductor (CMOS) platform.

Due to the dangling-bond-free two-dimensional (2D) structure of graphene, the direct growth of III-V compounds on graphene is in principle expected to be mediated by van der Waals forces, which relaxes the strict requirements of conventional heteroepitaxy, i. e., there are no constraints on either the surface symmetry or the lattice parameter of the overgrown material.[9–11] On the other hand, the lack of dangling bonds hinders the nucleation of III-V compounds on graphene.[12,13] Moreover, the formation of single-crystalline films over large areas is also impeded by the low surface energy of graphene which favors a 3D growth mode.[13] This phenomenon typically leads to the formation of bizarre nanostructures or highly defective layers.[4,13–19] In the case of GaN, the material of choice for solid-state lighting and high-frequency power electronics, the growth of smooth and single-crystalline layers has been demonstrated on epitaxial single-layer graphene synthesized on SiC,[11] where the periodic arrangement of step edges favors the nucleation of GaN.[11,20] The threading dislocation density of the GaN layers grown on graphene/SiC was, however, on the order of $1 \times 10^9$ cm$^{-2}$, a value as high as those obtained on AlN-buffered SiC and Al$_2$O$_3$ substrates.[21,22]

Nowadays, a common approach to avoid the formation of dislocations or to suppress their propagation along the growth direction consists in synthesizing the material in the form of nanowires (NWs) instead of films because the large surface to volume ratio of these nanostructures favors the elastic relaxation of strain and impedes the vertical propagation of extended defects.[23,24] Beside the reduced defect density, the NW geometry facilitates efficient light extraction[25] and enables the growth of non-polar[26] group-III nitride LEDs when fabricating core-shell heterostructures on the $M$-plane facets that constitute the NW sidewalls.[27] Furthermore, in comparison to planar devices, a decisive advantage of core-shell NW heterostructures is the possibility to drastically increase the total active area of the device for the same area of the wafer.[27] The synthesis of InAs, GaAs and InP NWs on graphene has been demonstrated using either molecular beam epitaxy (MBE) and/or metal-organic chemical vapor deposition (MOCVD).[12,28–30] Several groups have also recently demonstrated the growth of GaN NWs on graphite substrates as well as on different types of polycrystalline graphene layer structures transferred onto carrier substrates (Si, SiO$_2$ and Al$_2$O$_3$) by MOCVD and plasma-assisted molecular beam epitaxy (PA-MBE).[4,15,18,19,31–34] The GaN NWs synthesized by MOCVD, also known as nanorods, nucleate with a random orientation on graphene unless the latter is transferred onto a crystalline substrate.[19] Importantly, it has been conclusively shown that although the GaN MOCVD growth process might induce n-type doping and strain,[35] it does not significantly deteriorate the properties of graphene.[11,18,19,32] PA-MBE enables the growth of GaN NWs with typically smaller diameters than those attainable by MOCVD (down to 15–20 nm)[36,37] but it is not clear yet how the properties of graphene are affected by the growth of GaN. In fact, Hayashi et al.[34] used an AlN buffer layer to enhance GaN nucleation and to avoid the potential damage of graphene by the impinging active N species generated by the radio-frequency N$_2$ plasma source. Such an approach might protect graphene but the large series resistance introduced by the AlN layer is incompatible with the idea of using graphene as a built-in bottom contact for the fabrication of NW-based devices. Kumaresan et al.[33] have shown that GaN NWs can be epitaxially grown by PA-MBE on bare graphene but they did not analyze the properties of the remaining material after GaN growth. Regardless of the growth technique, GaN NWs prepared on transferred graphene exhibit a non-negligible tilt[4,33] (attributed to the formation of cracks, undulations and folds during the transfer of the graphene structures)[33] and are produced at the risk of contaminating the NWs as well as the growth equipment with residues derived from the transfer process. In addition, the transfer process complicates the fabrication of GaN/graphene hybrid structures on large-area wafers, as required for the mass production and commercialization of semiconductor devices.

In this letter, we investigate the self-assembled formation of GaN NWs by PA-MBE on several types of epitaxial graphene layer structures synthesized on SiC substrates. The different structures differ mainly in the total number of graphene layers that they contain. In contrast to transferred graphene, epitaxial graphene layer structures synthesized on SiC[38–41] are single-crystalline and can be directly used for the growth of GaN NWs, i. e., no processing is required upon the synthesis of graphene. It is thus an ideal system to investigate the nucleation of GaN on graphene as well as to elucidate how the PA-MBE growth process influ-ences the original graphene structure. Our study reveals that graphene is unstable against the active N species produced by radio-frequency $N_2$ plasma sources. During GaN growth, graphene is etched with a rate that is, however, several orders of magnitude lower than the GaN deposition rate. Hence, when the substrate contains a sufficiently large number of graphene layers, some of them withstand the growth process as demonstrated by Raman spectroscopy and transmission electron microscopy (TEM). The NWs grown on graphene nucleate preferentially at step edges, are single crystalline, vertical, well aligned, epitaxial, and of comparable structural quality as similar structures grown on conventional substrates. Therefore, our results demonstrate the feasibility of an *all-epitaxial* and easily scalable growth approach for the fabrication of GaN NW-based devices on graphene.

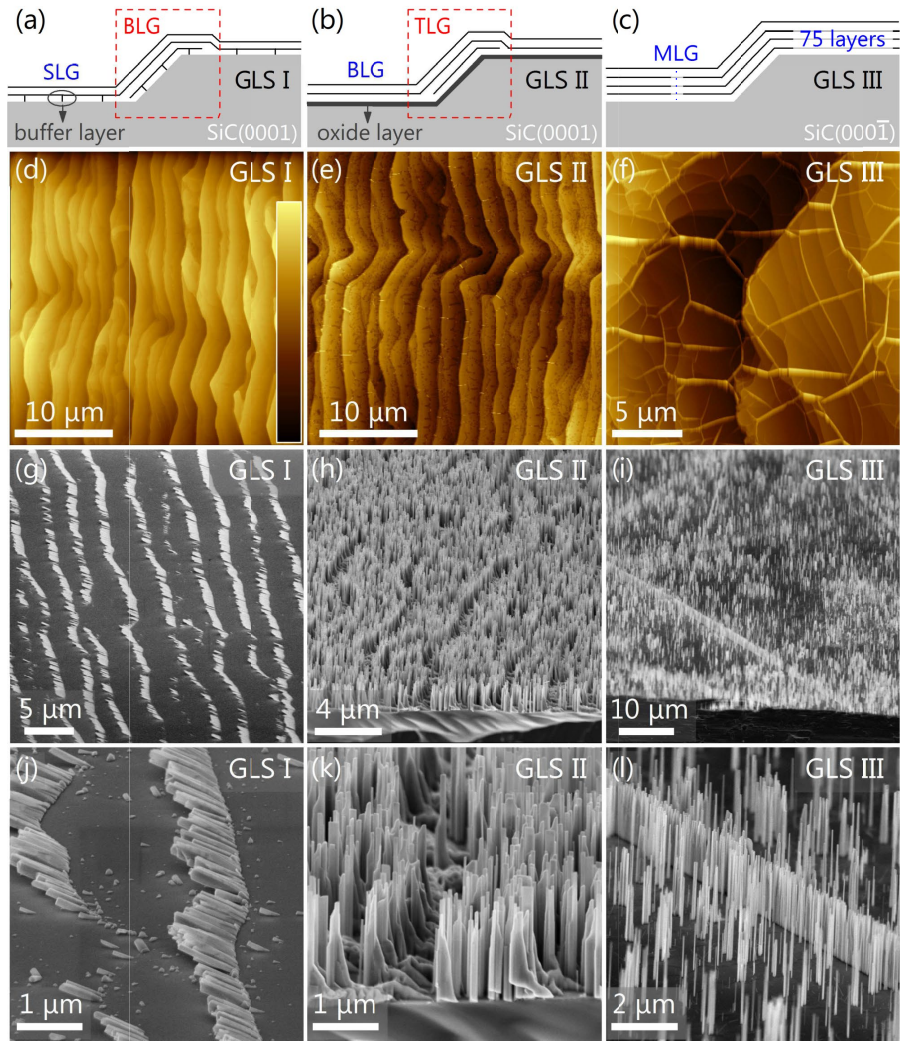

Figure 1: Sketches of the (a) Type I, (b) II, and (c) III epitaxial graphene layer structures (GLS) prepared on 6H-SiC substrates. SLG, BLG, TLG and MLG stand for singlelayer, bilayer, trilayer and multilayer graphene, respectively. (d–f) Atomic force micrographs of the corresponding pristine graphene layer structures. The linear color-code scale bar included in (d) represents 7, 15 and 60 nm in figures (d), (e) and (f), respectively. (g–l) Bird's eye view scanning electron micrographs of the GaN NW ensembles prepared on the 3 different types of graphene layer structures, as indicated in the images.

Figures 1(a–c) and 1(d–f) show schematic illustrations and atomic force micrographs, respectively, of the three types of graphene layer structures used in this work for the growth of GaN NWs. The Type I graphene layer structure (GLS I), schematically represented in Figure 1(a) and synthesized on SiC(0001), is characterized by the presence of approxiamtely 2 μm wide and flat terraces [Figure 1(d)] covered with single layer graphene (SLG). Graphene lies in this case on the top of the so-called buffer layer, which is a C layer isomorphic to graphene. The C atoms in this layer are arranged in a $6 \times (\sqrt{3} \times \sqrt{3})R30°$ lattice with respect to the underlying SiC substrate. The properties of the buffer layer differ from those of graphene because it exhibits covalent bonds to the substrate. The sample also contains inclusions of bi- to few-layer thick graphene at the step edges.[41–44] The Type II graphene layer structure (GLS II) is synthesized by annealing GLS I in air. During the annealing process, the buffer layer loses its bonding to the substrate due to the oxidation of the SiC substrate. The buffer layer is thus transformed into a new graphene layer.[41] As illustrated in Figure 1(b), this process leads to the formation of a bilayer graphene (BLG) film on the flat terraces and does not alter the overall surface morphology [see Figure 1(e)]. As in the case of the GLS I, a slightly larger number of graphene layers is expected nearby the step edges (probably trilayer graphene, TLG). The Type III graphene layer structure (GLS III) is synthesized on $SiC(000\bar{1})$ and contains $(75 \pm 15)$ layers of graphene.[45] We refer to such a structure as multilayer graphene (MLG). The graphene layers lie over both the terraces and the step edges of the SiC substrate. It is important to note that the properties of MLG differ from those of graphite because each C layer in the stack has the same electronic structure as an isolated graphene sheet.[43] As can be seen in the micrograph shown in Figure 1(f), the sample exhibits a prominent network of 5–20 nm high wrinkles. These wrinkles, common in epitaxial graphene structures formed on $SiC(000\bar{1})$, are likely a result of the different thermal expansion coefficients of SiC and graphene as well as the very weak coupling between these two materials.[43,45,46]

As a first step, we analyze the formation of GaN NWs on the GLS I. Figures 1(g) and 1(j) present bird's eye view scanning electron micrographs with different magnifications of the GaN NWs grown on this type of graphene layer structure. For the growth conditions employed in this work (see Methods section), NWs only form along the step edges, where the pristine substrate is covered by more than 1 layer of graphene [Figure 1(a)]. Hence, this type of substrate enables the formation of well-separated rows of GaN NWs. The NWs are about 1 μm in length, exhibit a diameter of approximately 100 nm, and nucleate very close to each other leading to the formation of coalesced aggregates. For lower substrate temperatures than the one used in this work, we also observe GaN nucleation on the terraces but in the form of a compact and faceted layer, as those commonly reported when similar growth conditions are used on AlN-buffered SiC(0001) and Si(111) substrates.[47–49] Interestingly, in the present case, the NWs formed at the step edges are highly tilted along a well-defined direction [see Figure 1(j)]. Since GaN NWs typically grow perpendicularly to the substrate normal,[37] this result suggests that here NWs nucleate exclusively on the few-layer thick graphene regions created on the semipolar facets of the SiC substrate [see Figure 1(a)].

Next, we investigate the growth of GaN NWs on the GLS II. Figures 1(h) and 1(k) present scanning electron micrographs of the resulting NW ensemble. In contrast to the previous type of substrate, NWs nucleate everywhere but at the step edges. The latter are clearly recognizable in Figures 1(h) and 1(k) by the periodic arrangement of approximately 0.5 μm wide stripes consisting of a 150–300 nm thick and faceted GaN layer. The NWs formed on the terraces, originally covered by BLG, are about 1.5 μm long, elongate along the substrate normal, and have diameters in the range of 25–90 nm. As for other types of substrates,[36,50] the NWs are closely spaced resulting in a significant degree of coalescence.

To analyze the impact of the PA-MBE growth process on the original GLS I and II, we employ μ-Raman spectroscopy. The Raman spectra before and after the growth of GaN NWs are shown in Figure 2(a) for the spectral range of the characteristic graphene 2D peak.[51] The spectra of the pristine GLS I and II exhibit the expected spectral positions and lineshapes of the 2D peaks for SLG and BLG, respectively.[44] Upon GaN growth, however, the Raman modes that reflect the existence of graphene are not detected anymore. Instead, a new Raman peak emerges in the spectral range between 2850 and 2950 cm$^{-1}$ in which C–H vibrational modes of hydrocarbons are commonly observed.[52] The C–H peak is absent in the spectra of the pristine graphene films and its intensity is correlated with that of the $E_2$ phonon peak from GaN NWs observed at 567 cm$^{-1}$ (not shown here). The origin of this peak will be further discussed below. These findings imply that SLG and BLG films do not withstand the growth of GaN NWs by PA-MBE.

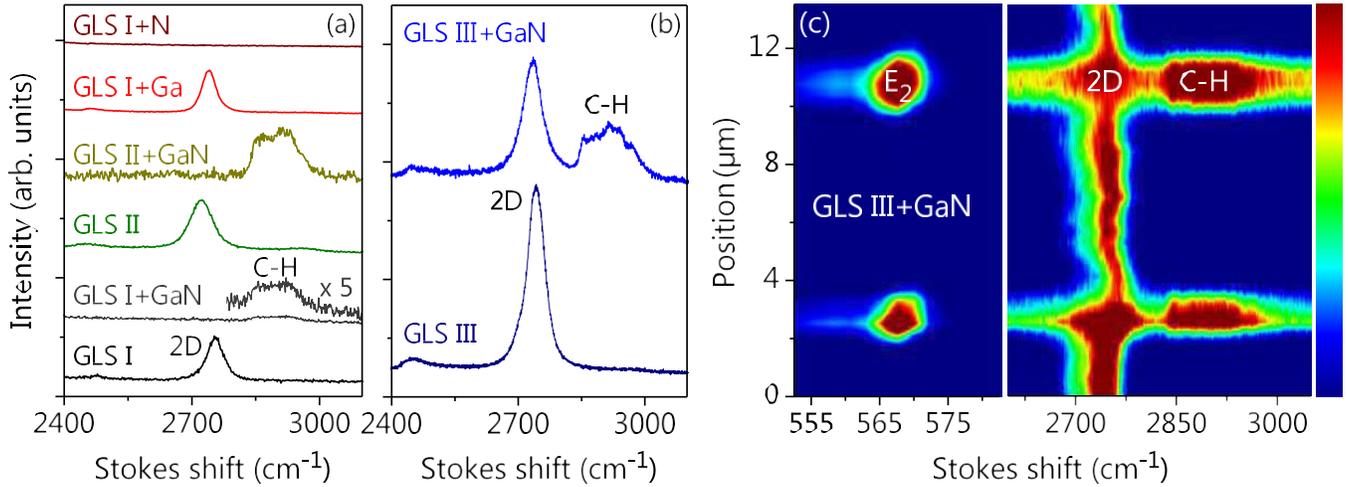

Figure 2: (a) Raman spectra from ensembles of GaN NWs prepared on the GLS I (GLS I + GaN) and II (GLS II + GaN). The spectra from the pristine GLS I and II as well as that of the GLS I after being exposed to either Ga (GLS I + Ga) or N (GLS I + N) at the GaN growth temperature are shown for comparison. (b) Raman spectra of the GLS III prior to and after GaN growth. The Raman spectra in (a) and (b) are vertically shifted for clarity. (c) Raman linescan for the sample containing GaN NWs on the GLS III (GLS III + GaN). The color-coded scale bar on the right indicates the intensity on a linear scale. The maximum intensity of the $E_2$, 2D and C–H peaks is saturated in this figure.

In order to elucidate the underlying reason for the degradation of the Type I and II graphene layer structures during GaN growth, we perform two control experiments where GLS I is exposed either to Ga or active N at the substrate temperature used for the growth of the GaN NWs. The Ga and N exposure times are the same as for the growth of the GaN NWs, i. e., 6 hours. The results of the analysis of these samples by μ-Raman spectroscopy are included in Figure 2(a). Upon exposing the surface to Ga, the graphene 2D peak remains unchanged. In striking contrast, after irradiating the surface with active N, we no longer detect the 2D peak. Therefore, the control experiments demonstrate that graphene is stable against Ga but is completely removed or chemically modified upon continuous exposure to active N. These results thus cast doubts upon the persistence of graphene in the experiments reported in Ref. 33, where the properties of the presumably remaining graphene layers were not analyzed after the growth of GaN NWs by PA-MBE.

Provided that the underlying reason for the degradation of graphene during N irradiation is the removal of C atoms by the impinging active N species, the use of substrates containing a larger number of graphene layers might alleviate or even solve the problem. Note that in order to provide electrical contact in between NWs, it is enough if a single layer of graphene withstands the growth process. Such a scenario motivates us to investigate the growth of GaN NWs on the GLS III shown in Figure 1(c), where a SiC($000\bar{1}$) substrate is covered with MLG. Figures 1(i) and 1(l) present bird's eye view scanning electron micrographs of the GaN NW ensemble prepared on this third type of graphene layer structure. We find that NWs nucleate preferentially at step edges as well as on structural defects. Several examples of nucleation at structural defects can be found in both micrographs as well as in that provided as Supporting Information. In particular, closely spaced NWs are seen to arrange in rows on scratches produced during the preparation of the original SiC($000\bar{1}$) substrate. The NWs that nucleate outside the scratches, i. e., on the regions covered by MLG, exhibit a much lower number density than those grown on the GLS I and II. In fact, we detect in this sample a large number of isolated NWs with rather small diameters (about 25 nm) despite of exhibiting a length of approximately 2 µm.

As for the previous substrates, the GLS III prior to and after GaN growth is analyzed by µ-Raman spectroscopy. As shown in Figure 2(b), the graphene 2D peak persists upon GaN growth and exhibits only slight changes with respect to the pristine GLS III. Apart from this peak, we also observe the C–H peak in this sample. We attribute the slight differences in the spectral position and lineshape of the 2D peak to minor structural modifications in the MLG film. This explanation is supported by the appearance of a relatively weak graphene D peak in the Raman spectrum at 1367 cm$^{-1}$ (see Supporting Information) which is known to be caused by the presence of structural defects.[53] Beside these changes, the 2D peak is also less intense after GaN growth. Nevertheless, judging from the relative intensity of this peak with respect to those of the SiC substrate, a significant number of graphene layers must be still present in the sample.

To elucidate whether there is MLG below and in between NWs, we record a spectral map from the MLG and GaN Raman signals along a linescan that crosses one complete terrace and two steps of the SiC substrate [Figure 2(c)]. A strong GaN Raman signal (the E$_2$ phonon line of wurtzite GaN at 568 cm$^{-1}$) is detected at the steps reflecting the large density of NWs at these positions (3 and 11 µm). The previous assignment of the C–H peak (2850–2950 cm$^{-1}$) to hydrocarbons adsorbed on the NW surfaces is proven here by the spatial coincidence of the observed Raman intensity enhancements. It is worth to mention that we have also observed the C–H peak in GaN NW ensembles prepared on other types of substrate like Si(111). Consequently, this peak is not related to the etching of the graphene layer structure but caused by the adsorption of hydrocarbons after GaN growth. Most importantly, the observation of the graphene 2D peak along the whole linescan provides evidence for the presence of a continuous MLG film. The persistence of MLG upon GaN growth renders this sample interesting and we thus investigate the structural and optical properties by x-ray diffraction (XRD), wet chemical etching, TEM, and low-temperature photoluminescence (PL) spectroscopy.

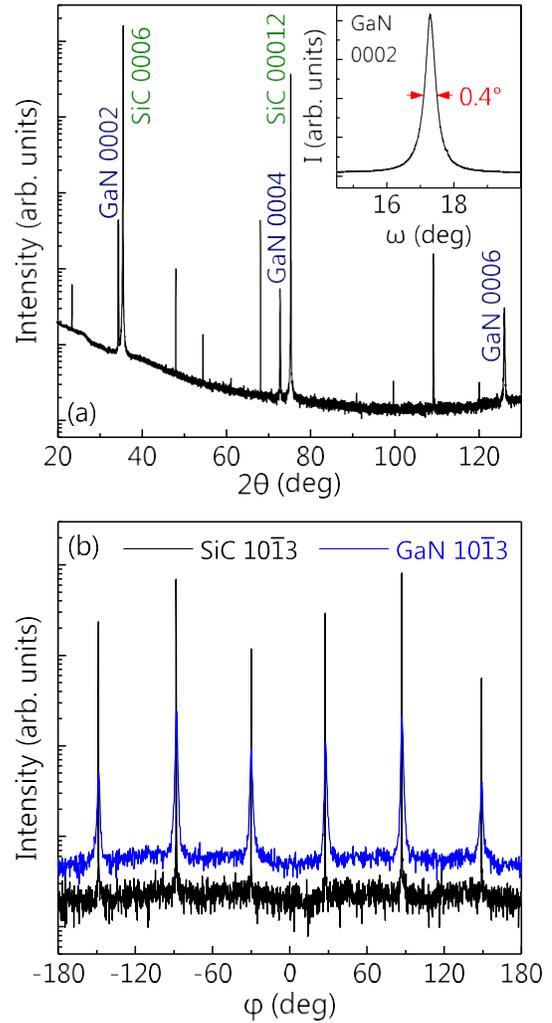

Figure 3: XRD analysis of the GaN NW ensemble prepared on the GLS III. (a) Symmetric θ/2θ scan. The narrow peaks without label stem from SiC forbidden reflections. The inset shows a $\omega$ scan across the symmetric GaN 0002 reflection. (b) $\phi$ scans across the SiC and GaN 10$\bar{1}$3 Bragg reflections.

To determine the epitaxial relationship between the GaN NWs and the remaining graphene layers, the sample is analyzed by XRD. Figure 3(a) shows a symmetric θ/2θ scan performed to assess the out-of-plane orientation of the GaN NWs. Apart from the peaks related to the 6H-SiC(000$\bar{1}$) substrate, we detect the 0002, 0004 and 0006 Bragg reflections from the GaN NWs. The lack of any additional GaN-related peak reveals that the NWs, which crystallize in the wurzite mod-ification, elongate along the (0001) axis, as further confirmed by TEM (see Supporting Information) and commonly reported for self-assembled GaN NWs grown by PA-MBE.[37,54] The out-of-plane orientation distribution of the GaN NWs, i. e., the tilt, is determined by a $\omega$ scan across the GaN 0002 Bragg reflection. As shown in the inset of Figure 3(a), the NWs exhibit a tilt distribution of only 0.4°. This value is more than four times lower than those reported in Refs. 4 and 33 for GaN NW ensembles synthesized on transferred graphene. The broad out-of-plane orientation distribution observed in the latter case was attributed to an imperfect transfer process of the graphene layer structure.[33] The in-plane orientation of the NWs with respect to the graphene layers and the SiC substrate is investigated here by measuring $\phi$ scans across the GaN and SiC $\{10\bar{1}3\}$ Bragg reflections in skew geometry [Figure 3(b)]. In both cases, we detect six peaks at identical $\phi$ values that reflect the six-fold symmetry of the hexagonal GaN and SiC crystal lattices. The GaN NWs are thus in registry with the substrate in such a way that SiC $\langle 1\bar{1}00 \rangle \parallel$ GaN $\langle 1\bar{1}00 \rangle$. Since it is known that the zigzag edges of the graphene layers are perpendicular to the SiC $\langle 1\bar{1}00 \rangle$ directions,[38] same applies to the relationship between the GaN and graphene lattices. This epitaxial relationship is the same as the one found in Ref. 33 for GaN NWs grown by PA-MBE on graphene layers transferred onto amorphous SiO$_2$.

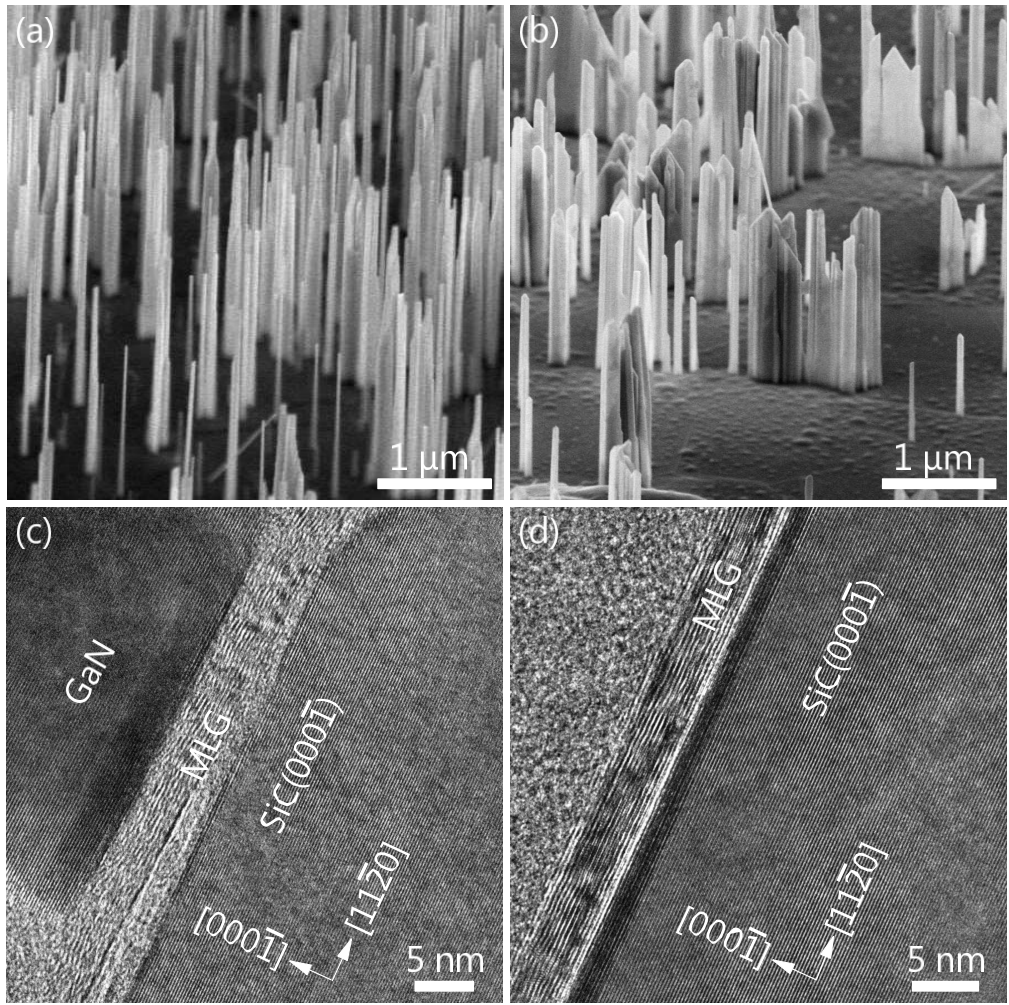

Figure 4: Bird's eye view scanning electron micrographs of the GaN NWs prepared on the GLS III (a) prior to and (b) after KOH etch. (c,d) Bright-field cross-sectional high-resolution transmission electron micrographs of the sample where GaN NWs are grown on the GLS III. Figure (c) is acquired at the bottom part of a NW and Figure (d) in a region in between NWs. The thickness of the remaining graphene structure seen in (c) and (d) is 5.5 and 4.8 nm, respectively.

Our XRD experiments demonstrate that the NWs grown on MLG are epitaxial and elongate along the polar ⟨0001⟩ axis. To clarify whether the growth direction is the [0001] or the [000$\bar{1}$] one (known as Ga- and N-polar orientations, respectively), which has important implications for the fabrication of GaN NW-based devices,[26,55] the sample is investigated by KOH chemical etching. As discussed in detail in Refs. 56, 57 and 26, KOH only etches GaN NWs when they grow along the [000$\bar{1}$] direction, i. e., when they are N polar. Figures 4(a) and 4(b) present bird's eye view scanning electron micrographs of the sample prior and after being exposed to KOH, respectively. The micrographs reveal that the NWs are etched and develop a pencil-like shape when they are exposed to KOH. This result demonstrates that GaN NWs grown on graphene are N polar, as typically observed for self-assembled GaN NWs grown by PA-MBE on a wide variety of substrates.[26,49]

The MLG structure after GaN growth is analyzed in the following on a microscopic scale by high-resolution TEM. Representative bright-field transmission electron micrographs of the graphene layers of regions below and in between NWs are shown in Figures 4(c) and 4(d), respectively. In both micrographs, we detect the presence of a continous MLG film. The micrographs also reveal the absence of any additional compound on the surface as well as a significant reduction in the thickness of the MLG structure as compared to its original one of (75±15) layers. After GaN growth, the thickness of the MLG structure is found to be about 5 nm (≈15 layers). Hence, during the formation of the NWs several tens of layers are etched by the impinging active N species. The remaining graphene layers are partly distorted underneath the NWs [Figure 4(c)] and perfectly smooth and continuous in between them [Figure 4(d)]. The amorphous interlayer observed in the left bottom part of Figure 4(c) is attributed to a detachment of the MLG film occurring during the preparation of the samples for TEM. This detachment of the graphene layer structure is also observed in the regions in between NWs (see Supporting Information). The results derived from the analysis of the samples by HR-TEM allow us to conclude that the exposure of graphene to active N does not result in the formation of a different compound but in an etching of graphene. Taking into account the number of graphene layers left, for the experimental set-up and growth conditions used in this work, the etch rate of graphene by an active N flux of 0.5 ML/s is on the order of 3 × $10^{-3}$ layers/s.

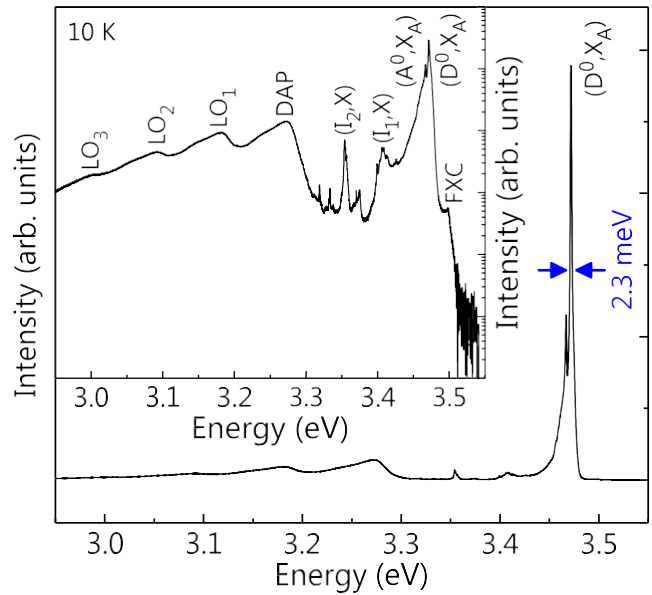

Figure 5: Low-temperature (10 K) cw PL spectrum of the GaN NW ensemble prepared on the GLS III. The inset shows the same spectrum but on a logarithmic scale.

Finally, the structural perfection of the GaN NWs is analyzed by low-temperature (10 K) continuous-wave (cw) PL. Figure 5 shows the near band-edge μ-PL spectrum of the GaN NWs (we excite about several tens of NWs) on a linear scale. The spectrum was acquired in a region free of scratches (see also Supporting Information). As can be seen in Figure 5, the PL spectrum is dominated by the donor-bound exciton transition ($D^0,X_A$) at 3.471 eV, as expected for a GaN crystal free of homogeneous strain.[58] The linewidth of this transition, which depends on the residual inhomogeneous strain introduced by structural defects[59] as well as on the NW diameter distribution due to the varying distances of the unintentionally incorporated O donors to the NW sidewall surfaces,[60–62] is 2.3 meV. This value is comparable to those observed in GaN NW ensembles grown on Si(111) (1–3 meV)[58–61,63,64] and significantly lower than the 14 meV reported in Ref. 33 for GaN NWs prepared by PA-MBE on graphene layers transferred onto amorphous $SiO_2$. For a closer look at all transitions observed in the PL spectrum, it is plotted on a logarithmic scale in the inset of Figure 5. We clearly distinguish several additional transitions: (i) at 3.498 eV, the intrinsic free exciton C transition (FXC), (ii) at 3.467 eV, an acceptor bound exciton transition ($A^0X_A$) attributed to unintentional incorporation of Mg[65,66] (frequently used as p-type dopant in our PA-MBE system), (iii) centered at 3.401 and 3.355 eV, two bands labeled as ($I_1$,X) and ($I_2$,X) which are associated with the recombination of excitons bound to $I_1$ and $I_2$ basal-plane stacking faults (SFs),[67] respectively, and (iv) at 3.272 eV, a donor-acceptor pair transition (DAP), together with its longitudinal optical (LO) phonon replicas at 3.18, 3.09 and 3 eV, resulting from the residual doping by O and Mg.[65,66] The intensity ratio between the ($D^0$,$X_A$) and the ($I_{1,2}$,X) transitions is ≈ 40. Since SFs in GaN NWs act as quantum wells confining excitons that recombine purely radiatively at 10 K,[68] a very low density of SFs (namely, on the order of 1 SF per several tens of NWs) is sufficient to explain the observed intensity ratio, as shown elsewhere.[69] Note that, we do not observe any pronounced emission related to inversion domain boundaries at 3.45 eV.[70,71] Furthermore, we do not detect any yellow luminescence when the PL measurements are extended towards longer wavelengths (not shown here). The structural perfection of these GaN NWs is thus absolutely comparable to that of standard GaN NW ensembles grown on Si(111) by PA-MBE.[58]

To summarize and conclude, we have analyzed the self-assembled growth of GaN NWs by PA-MBE on several types of epitaxial graphene layer structures prepared on SiC substrates. Our study demonstrates that graphene is etched during GaN growth by the impinging active N species. In view of the fact that the concentration of ionized active N species in the plasma is known to depend on the type of radio frequency plasma-source as well as on the operation conditions (the power applied and the $N_2$ flow), it might be possible to minimize or even suppress the etching of graphene during the formation of GaN NWs in PA-MBE. Further studies are required to elucidate this issue that might affect the conclusions of previous reports on the growth of GaN NWs on graphene.[33] Moreover, these results are not only relevant for the GaN NW commnunity but also of interest for the fabrication of graphene/BN heterostructures[72] by PA-MBE. Regardless this possibility, graphene is etched very slowly ($\approx 3 \times 10^{-3}$ layers/s) for the present conditions and several graphene layers survive the growth process when employing a MLG film for the synthesis of GaN NWs. We find that the NWs, which preferentially nucleate at step edges and morphological defects, are interconnected through the remaining graphene layers as desired for the fabrication of GaN NW-based devices on graphene. The NWs are vertical, elongate along the [$000\bar{1}$] direction, and exhibit a well-defined in-plane epitaxial relationship with the substrate, namely, the GaN $1\bar{1}00$ directions are perpendicular to the zigzag edges of the graphene layers. The structural perfection of these NWs is comparable to those of standard GaN NW ensembles prepared by PA-MBE on conventional substrates as demonstrated by PL spectroscopy. In comparison with similar structures grown by PA-MBE on transferred graphene, the NWs prepared on epitaxial graphene are better oriented and exhibit narrower excitonic transitions reflecting a reduced concentration of structural defects. Consequently, the synthesis of GaN NWs on epitaxial graphene does not only enable the fabrication of hybrid GaN/graphene devices on large area wafers without the technical inconveniences inherent to the graphene transfer process, but also improves the morphological and structural properties of the final NW ensembles.

After demonstrating here the synthesis of GaN NWs on epitaxial MLG films, it would be highly interesting to analyze the electrical properties of the resulting GaN/graphene interface as a next step. Such a study could then be used to conclusively evaluate the actual potential of using epitaxial graphene as a built-in bottom contact for the fabrication of NW-based devices by PA-MBE.

## Methods

### Epitaxial growth of graphene on 6H-SiC substrates

Epitaxial graphene is synthesized on 6H-SiC(0001) and ($000\bar{1}$) substrates by employing the surface graphitization method in an inductively heated furnace.[42] Prior to graphene synthesis, the substrates are cleaned in n-butylacetate, acetone, and ethanol, followed by a hydrogen etching treatment carried out at 1400 °C for 15 min in a forming gas atmosphere (atomic concentrations: 95 % Ar and 5 % H) of 900 mbar and using a flow rate of 500 sccm. The latter is performed in the same furnace where graphene is grown. The GLS I is prepared on SiC(0001) by thermally treating the samples at 1600 °C for 15 min in an Ar atmosphere of 900 mbar and employing a flow rate of 500 sccm.[42] The GLS II is produced via oxygen intercalation by thermally treating the GLS I at 600 °C for 40 min in air using a resistively heated furnace.[41] Finally, the GLS III is synthesized on SiC(000$\bar{1}$) at 1450 °C for 60 min in high-vacuum conditions (around $10^{-5}$ mbar) using a confinement-controlled-sublimation environment.[43,45]

## Growth of GaN NWs by PA-MBE

After the formation of the graphene layer structures, the backside of the substrates is coated with an approximately 1 μm thick Ti layer for efficient thermal radiation coupling during the subsequent growth of the GaN NWs by PA-MBE. The NWs are grown in a PA-MBE system equipped with a solid source effusion cell for Ga and an SVT RF 4.53 radio-frequency N plasma-source to generate active N. The radio-frequency N plasma-source is operated at 500 W using a $N_2$ flow of 3.5 sccm. To guarantee N-rich growth conditions, as required for the self-assembled formation of GaN NWs in PA-MBE,[36,73] the Ga and N fluxes employed for the growth of the GaN NWs are set to 0.3 and 0.5 ML/s, respectively (a growth rate of 1 ML/s is equivalent to $1.14 \times 10^{15}$ atoms cm$^{-2}$ s$^{-1}$). The substrate temperature during the growth of the GaN NWs is 1060 °C, as measured with a thermocouple placed on the heater. We estimate that the actual substrate temperature is, however, approximately 250–300 °C lower than this value. To initiate the growth, upon igniting the radio-frequency N plasma-source, the Ga and N shutters are opened simultaneously. The total growth time is about 6 hours.

## Characterization

The morphological and structural properties of the samples are investigated by scanning electron microscopy, XRD, TEM and wet chemical etching. Scanning electron micrographs are acquired using a field-emission microscope. XRD experiments are performed with CuK$_{\alpha 1}$ radiation using a Panalytical X-Pert Pro MRD system equipped with a Ge(220) hybrid monocromator. All measurements are recorded with a 1 mm slit in front of the detector. Cross-sectional TEM specimens are prepared using the standard method of mechanical grinding and dimpling down to below 25 μm followed by Ar-ion milling. Transmission electron micrographs are acquired using a JEOL JEM 3010 microscope operating at 200 kV and equipped with a Gatan charge coupled device camera. To assess the polarity by chemical etching,[26] the samples are exposed to a 5 M KOH aqueous solution at 40 °C for 10 min.[57]

The optical properties of the samples are analyzed by Raman, cw PL, and cathodoluminescence (CL) spectroscopy. Raman measurements are performed at room temperature in backscattering configuration. The 473 nm line of a solid-state laser is focused onto the sample by a confocal microscope (spatial resolution of 1 μm). The scattered Raman signal is collected by the same objective, dispersed spectrally by a 80-cm Jobin-Yvon monochromator equipped with a grating with 600 lines/mm, and detected by a LN$_2$-cooled charge-coupled device (CCD) camera. PL experiments are performed at 10 K with the 325 nm line of a HeCd laser. The PL signal is dispersed by a spectrometer with 2400 lines/mm and detected by a CCD camera. The laser beam is focused to a 1 μm spot and the excitation density is 10 W cm$^{-2}$. Imaging and spatially resolved luminescence measurements (shown in the Supporting Information) are recorded using a scanning electron microscope with an attached CL system. The system is equipped with a field-emission electron gun and a He-cooling stage allowing the sample temperature to be controlled in the range from 7 to 300 K. A photomultiplier tube is used for the acquisition of monochromatic images and a CCD for recording CL spectra. Throughout the experiments, the acceleration voltage and the probe cur-rent of the electron beam are set to 5 kV and 0.75 nA, respectively. Samples are cooled down to 10 K for the acquisition of CL spectra. The spectral resolution amounts to 1 meV.

# Supporting information

Spectrally extended µ-Raman spectra of the GLS III prior to and after GaN growth; complementary high-resolution transmission electron micrographs of the MLG film in between GaN NWs; and additional transmission and scanning electron micrographs as well as low-temperature cathodoluminescence spectra of the GaN NWs grown on the GLS III.

# Acknowledgements

We thank Carsten Stemmler for his help to prepare the samples and his dedicated maintenance of the MBE system together with Hans-Peter Schönherr and Claudia Herrmann. We are indebted to Anne-Kathrin Bluhm for her technical assistance to acquire scanning electron micrographs. We also thank Martin Heilmann for fruitful discussions and a critical reading of our manuscript. Financial support provided by the Leibniz-Gemeinschaft under Grant SAW-2013-PDI-2 is gratefully acknowledged. P.C. acknowledges funding from the Fonds National Suisse de la Reserche Scientifique through project No. 161032, and Z.d.S.S. funding from FAPESP under grant 2013/256253.

**Table of Contents Graphic**

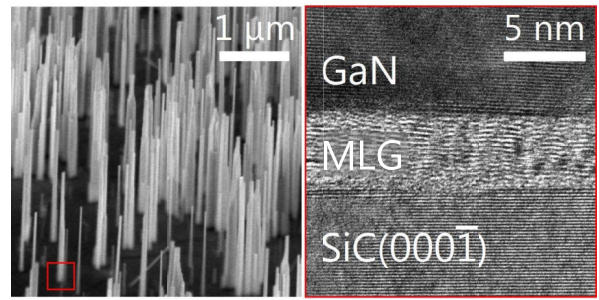